\documentclass[12pt]{article}
\pdfoutput=1
\usepackage[numbers,sort&compress]{natbib}
\usepackage{graphicx}
\usepackage{amsmath,graphicx,amssymb,subfigure}
\usepackage[usenames,dvipsnames]{color}
\usepackage{amssymb}
\usepackage{amsmath}
\usepackage{isomath}
\usepackage{amsthm}
\usepackage{enumerate}
\usepackage{stmaryrd}
\usepackage{amsfonts}
\usepackage[all]{xy}
\usepackage[section]{placeins}
\usepackage[colorlinks=true,linktocpage=true,linkcolor=blue,citecolor=blue]{hyperref}

\newcommand{\be}{\begin{equation}}
\newcommand{\ee}{\end{equation}}
\newcommand{\bea}{\begin{eqnarray}}
\newcommand{\eea}{\end{eqnarray}}

\begin{document}

\bibliographystyle{hieeetr}

\pagestyle{plain}
\setcounter{page}{1}

\begin{titlepage}

\begin{center}

\vskip 30mm

{\Large\bf Quantum computational complexity, Einstein's equations and accelerated expansion of the Universe}

\vskip 0.8cm

Xian-Hui Ge $^{~1}$,~~~  Bin Wang  $^{~2,3}$ \\
$^1${\it Department of Physics, Shanghai University, Shanghai 200444,  China \\
and Department of Physics, University of California at San Diego, CA92097, USA}\\
$^3${\it Center for Gravitation and Cosmology, College of Physical Science and Technology,
Yangzhou University, Yangzhou 225009, China}\\
$^4${\it Department of Physics and Astronomy, Shanghai Jiao Tong University, Shanghai, 200240, China}\\
\medskip

\vspace{5mm}
\vspace{5mm}

\begin{abstract}
We study the relation between quantum computational complexity and general relativity. The quantum computational complexity is proposed to be quantified by the shortest length of  geodesic quantum curves. We examine the complexity/volume duality in a geodesic causal ball in the framework of Fermi normal coordinates and derive the full non-linear Einstein equation. Using insights from the complexity/action duality, we argue that the accelerated expansion of the universe could be driven by the  quantum complexity  and free from  coincidence and fine-tunning  problems.
\end{abstract}
\end{center}
\noindent
\end{titlepage}

\section{Introduction}
Recent years, great efforts have been devoted to understanding the deep connections between fundamental concepts of quantum information theory and spacetime geometry. In the context of the AdS/CFT correspondence, Ryu and Takayanagi proposed that entanglement entropy of a subsystem, the  measurement of degrees of freedom between subsets in general quantum states, corresponds to the area of the minimal bulk  surface at the boundary of this subregion \cite{RT1,RT2,taka}. This connection was further studied to relate the first law of entanglement in the vacuum of the boundary CFT to Einstein's equations linearized around the AdS vacuum in the bulk \cite{liei,swingle12,van16,czech, czech2,van}. Jacobson further derived the full non-linear Einstein equation in a small geodesic ball under the extremal vacuum entanglement entropy hypothesis \cite{jacobson16} (see also \cite{casini, czech3,jacobson95,GB,antony} for further reading).

The recent development of  ``complexity=action" (CA) and ``complexity=volume" (CV) conjectures are suggested as new entries in the holographic dictionary \cite{susskind,susskind15, susskind16,susskind17,rath}. The CA conjecture states that the quantum complexity of the boundary state equals to the gravitational action evaluated on a bulk region known as the Wheeler-DeWitt patch, while the CV conjecture identifies the complexity of the boundary state with the spatial volume of a maximal slice behind the horizon. Both conjectures will be explored in this work. Quantum computational complexity $\mathcal{C}$ (in brief,  \textit{quantum complexity}) is defined as the minimum number of elementary operations needed to produce the target state of interest from a reference state. Originated from the field of quantum computations, quantum complexity grows linearly in time under the evolution of a local Hamiltonian and its growth rate is then proportional to the number of active degrees of freedom. The definition and calculation of quantum complexity in quantum many body systems were investigated in recent works \cite{chapman17, mayers17}.

Considering that spacetime geometry can be represented by the entanglement structure of the underlying microscopic quantum states, in this work we are going to further investigate the relation between quantum complexity and the evolution of our Universe.  One of the most mysterious problems in  cosmology is that
$95\%$  components of our Universe need to be explained properly \cite{Planck16}. This motivates us to reconsider the starting point about our theory of spacetime and cosmology.

 Assuming that the CA and CV dualities have general applicability, it would be of great interest to discuss their physical interpretations in cosmology.
 We will first build the connections between quantum complexity and geodesic quantum distance by scrutinizing quantum Fisher  information  metric in a Hilbert space.  The quantum complexity is then defined by the minimal length measured by the geodesic quantum distance from a reference sate to a target state.  We then examine the CV conjecture in a geodesic causal ball in the framework of Fermi normal coordinates and derive the full non-linear Einstein equation, in particular cosmological Friedmann equations. The derivation is valid under the condition that the radius of the ball is much smaller then the local curvature length, but this limitation can be overcome in terms of conformal Fermi coordinates.
 The emergence of the Einstein equation from the CV duality implies that the underlying microscopic degrees of freedom of our universe is somehow linked to quantum complexity.  Considering large scale structure of the Universe and further examining the CA duality in the cosmological setup,  we are able to find some hints that the accelerated expansion of our universe may be driven by the quantum complexity.

\section{ Quantum complexity and geodesic quantum distance}
Without loss of generality, we  consider a family of parameter-dependent Hamiltonian $H(\lambda)$ requiring a smooth dependence on a set of parameters $\lambda=(\lambda_1,\lambda_2,\cdots)\in \mathcal{M}$, which consists of the base manifold of the quantum system. The Hamiltonian acts on the parameterized Hilbert space $H(\lambda)$ with
$|\varphi_{n}(\lambda)\rangle$ denoting the eigenstates. Suppose there is a system state $|\psi(\lambda)\rangle$, which is a linear combination of $|\varphi_{n}(\lambda)\rangle$ at each point in $\mathcal{M}$. For a reference state $|\psi_R\rangle$, which could be the ground state of the system,  its relation to the target state can be described by  $|\psi_T\rangle=U|\psi_R\rangle$.
The complexity of the unitary operator $U$ is associated with the minimal number of gates necessary to approach $U$. It was proved in \cite{nielsen} that the minimal geodesic between the identity operation and $U$ is essentially equivalent to the number of gates required to synthesize $U$. So,  in principle, quantum complexity $\mathcal{C}$ can be defined as the shortest path between two points in the manifold $\mathcal{M}$.

Upon infinitesimal variation of the parameter $d\lambda$, the quantum distance between $|\psi(\lambda+d\lambda)\rangle$ and $|\psi(\lambda)\rangle$ can be measured by the quantum fidelity \cite{QI}.  Up to the second order in $\delta \lambda$, the fidelity read
\be\label{fidelity}
F=\big|\langle \psi(\lambda) | \psi(\lambda+d \lambda)\rangle\big|=1-\frac{1}{2}\mathcal{G}_{ab}\delta \lambda^a \delta \lambda^b+\mathcal{O}(\delta \lambda)^3,
\ee
where $\mathcal{G}_{ab}$ is the \textit{quantum Fisher-Rao information metric }on the manifold $\mathcal{M}$ of probability distributions. The Fisher-Rao information metric can also be generated by the relative entropy (see Appendix B). In general, the real part of the Fubini-Study metric reduces to the Fisher-Rao information metric. It was shown in \cite{CM} that the Fisher information metric is invariant under reparametrization of the sample space  and  it is covariant under reparametrizations of the manifold, i.e. the parameter space, see e.g. \cite{covariantM} for a review. The discussions of quantum fidelity in curved spacetime can be found in \cite{ge06,ge08}.

The geodesic quantum distance between two quantum states $\psi (\lambda_I)$ and $\psi (\lambda_F)$ measured by the Fisher-Rao metric $\mathcal{G}_{ab}$ in quantum information theory is then given by \be \mathcal{S}=\int^{\lambda_F}_{\lambda_I}\sqrt{|\mathcal{G}_{ab}d\lambda^{a}d\lambda^{b}|},\ee
where $\mathcal{S}$ can also be interpreted as the length of the geodesic curve. The Fisher-Rao metric $\mathcal{G}_{ab}$ measures the geodesic distance of points lying on the Bloch sphere since the inner product of any two quantum states should be within the range $[0,1]$. One can thus define $|\langle \psi | \chi\rangle|\equiv \cos^2\frac{\theta}{2}$, so that
\be
d\theta=2 d\mathcal{S}=2\sqrt{|\mathcal{G}_{ab}d\lambda^{a}d\lambda^{b}|}.
\ee
A remarkable feature of the Fisher-Rao metric as a distinguishability measurement is characterized by its appearance in the time-energy uncertainty relation.
For a non-adiabatic system, let us label the evolution of the state  by time $t$ rather than the parameter $\lambda$. Expand $|\psi(t+dt)\rangle$ to the second order in $dt$
\be
| \psi(t+dt)\rangle=| \psi(t)\rangle+\frac{d}{dt}| \psi(t)\rangle dt+\frac{1}{2}\frac{d^2}{dt^2}| \psi(t)\rangle dt^2+...
\ee
With the help of Schr$\ddot{o}$dinger's equation and after some manipulations, we arrive at
\be
\big|\langle \psi(\lambda) | \psi(\lambda+d \lambda)\rangle\big|=1-\frac{1}{2}\frac{(\Delta  E)^2}{\hbar^2}dt^2+\mathcal{O}(dt^4).
\ee
Comparing with the definition of the Fisher-Rao metric (\ref{fidelity}), we obtain
\be\label{veloc}
\frac{d\theta}{dt}=2\frac{d\mathcal{S}}{dt}=\frac{2\Delta E}{\hbar}.
\ee
The term ${d\mathcal{S}}/{dt}$ denotes as the quantum velocity.
This is indeed a precise version of time-energy uncertainty relation: the system evolves quickly through regions where the uncertainty in energy is large.

To connect the above discussions to the notion of quantum complexity, one may notice  that the growth of complexity is conjectured to be bound by the energy as \cite{susskind15}
\be\label{complex1}
\frac{d\mathcal{C}}{dt}\leq \frac{2\Delta E}{\pi\hbar},
\ee
where $\Delta E$ is the energy difference between $\psi(\lambda+d \lambda)$ and $\psi(\lambda)$.  Comparisons between (\ref{veloc}) and (\ref{complex1}) suggest complexity indeed relate to the geodesic quantum distance as $ \mathcal{C} \sim  \mathcal{S}$. One can further define the complexity $\mathcal{C}$ as the shortest length of the geodesic curve from a reference state to a target state as in \cite{chapman17},  \be\label{cd}
\mathcal{C} \equiv  \min \mathcal{S}(| \psi\rangle).
\ee
Up to now, both $\mathcal{C}$ and $\mathcal{S}$ are evaluated on the dual field theory side. In calculations in gravity, $\Delta E$ will be related to the energy in the bulk theory.
\section{ The complexity/volume duality  and the Einstein equation}
\label{section3}
\begin{figure}
\begin{center}
	\includegraphics[height=7cm,trim={2.5cm  20.8cm 2cm 3.5cm},clip]{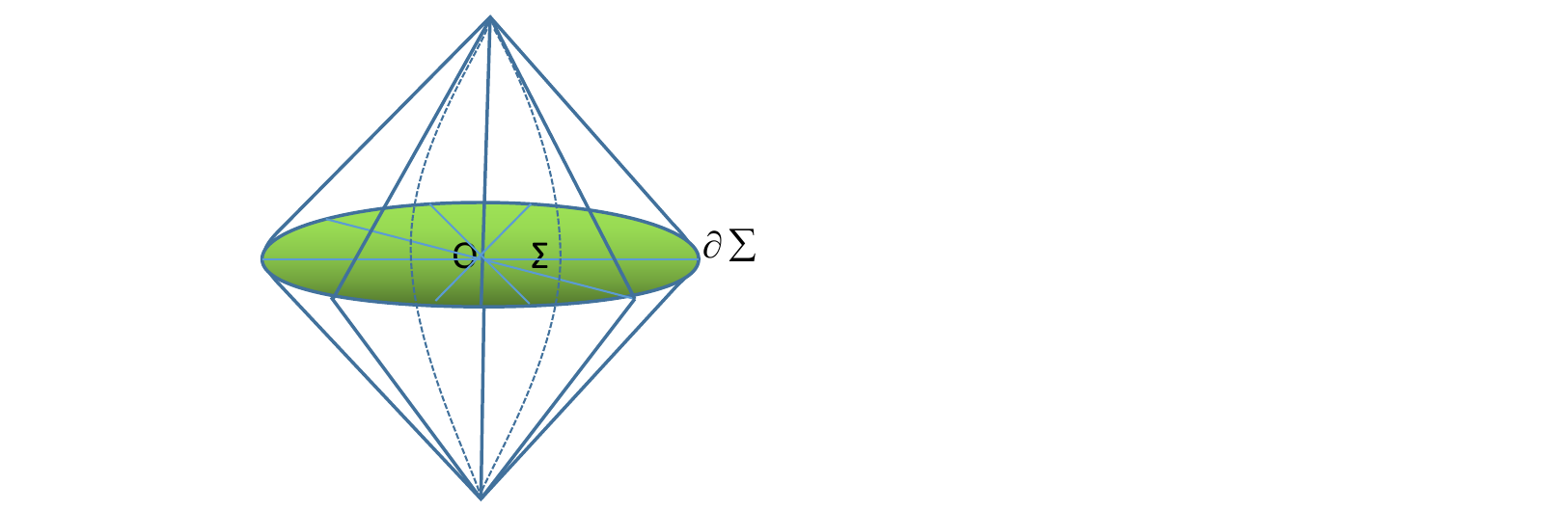}
\end{center}
	\caption{Causal diamond of a ball-shaped region $\Sigma$ of radius $l$ with center $O$ and boundary $\partial \Sigma$. The FNC system is central at the geodesic $O$ of a free observer. The construction of the causal diamond is valid at an arbitrary time.   }\label{diamond}
\end{figure}
Our logic to derive the Einstein equation is as follows: within an infinitesimal time period $\delta t$, we assume an infinitesimal variation of the quantum state $d\lambda$ and the resulting growth of the complexity is dual to the volume deficit evaluated in Fermi normal coordinates. The Einstein equation and also the Friedmann equation hence emerge as a consequence of the CV duality.
The CV duality states that the complexity of the boundary state is proportional to the volume of a maximal bulk surface and asymptotes to the time slice on $\Sigma$ on which the boundary state is defined
\be\label{cvduality}
\mathcal{C}(\Delta t) \sim \frac{V (\Delta t)}{G l},
\ee
where $l$ is some length scale associated with the bulk geometry and $G$ is the Newton constant. This volume is bounded by the spatial slices at times $t_1$ and $t_2$ ($\Delta t=t_2-t_1$) on the boundaries.  As small perturbations imposed on the quantum system, quantum complexity and also entanglement entropy grow with time. In turn, the spacetime geometry (Einstein-Rosen Bridge) is also blown up \cite{ms2013}.

 Let us first estimate the right hand of (\ref{cvduality}).  As sketched in Fig.\ref{diamond}, we  consider the Fermi normal coordinates (FNC)  system central at the geodesic $O$ in a spacetime of dimension $d$ (see Appendix B for introductions on Fermi normal coordinates). The geodesics sending out from $O$ orthogonal to $u^a$ forms a
$(d-1)-$dimensional spacelike ball $\Sigma$ with the ball radius $l$. Consider a FNC system based at $O$, with the timelike coordinate  $x^0_{F}$ and spacelike ones $x^i_{F}=r n^i$, where $r$ is the geodesic distance and $n^i$ is a unit vector at $O$ yielding $\delta_{ij}n^in^j=1$. We assume that the radius of the ball is much smaller than the local curvature length. For cosmology, this corresponds to $l\ll H^{-1}$ with $H$ the cosmological Hubble paramter. Note that this restriction can be relaxed by considering conformal Fermi coordinates.

The volume element of $\Sigma$ to the second order of the FNC coordinates is given by
\be
d V=\sqrt{h}d^{d-1}x=\bigg(1-\frac{1}{6}r^2~ {}^FR_{ik}^{~~i}{}_{l}n^k n^l\bigg)r^{d-2}drd\Omega,
\ee
where $h$ is the determinant of the spatial metric $h_{ij}$ on $\Sigma$,  ${}^FR_{ikj}^{~~i}$ is the spatial Ricci scalar at $O$  and $d\Omega$ denotes the area element on the unit $(d-2)$-sphere. For spherical symmetry,  intergrating over $r$ from $0$ to $l$ yields
\be
V=l^{d-1}\Omega_{d-2}-\frac{\Omega_{d-2}l^{d+1}}{6(d-1)(d+1)}\mathcal{R},
\ee
where  $\mathcal{R}={}^FR_{il}^{~~il}$ is the spatial Ricci scalar at $O$ and we have used the integrand $\int d\Omega n^k n^l=\frac{\Omega_{d-2}}{d-1}\delta^{kl}$.
The volume variation per unit time compared to Minkowski space at $\mathcal{R}=0$ is then given to the lowest order
\be\label{variation}
\frac{\delta V}{\delta t}=V(\mathcal{R}\neq 0)-V(\mathcal{R}= 0)=-\frac{\Omega_{d-2}l^{d+1}}{6(d-1)(d+1)}\mathcal{R}.
\ee
According to the CV duality and (\ref{cd}), the complexity growth per unit time  equals to the maximal volume variation per unit time
\be
\frac{\delta \mathcal{C}}{\delta t} \sim \frac{|\delta V| }{G l \delta t}\sim \Delta E.
\ee

Now we are going to evaluate $\Delta E$. For a conformal field theory, $ E$ equals to $H_{\zeta}$ since the modular Hamiltonian generates the flow of the conformal boost Killing vector \cite{jacobson16}.
But for a general quantum field theory, a general state and a general region, the modular Hamiltonian is not known and there is no available  practical method to compute it.
The generator of this flow in the underlying CFT may be written covariantly as
\be\label{hz}
H_{\zeta}=\int_{\Sigma} T^{ab}\zeta_b d\Sigma_a.
\ee
Choose the radius of the geodesic ball to be much smaller than any length scale in the geometry, but still much larger than the Planck scale $\ell_p$ so that the quantum gravity effects can be neglected. In this small ball limit, the energy density can be treated as a constant throughout this region and the variation of $H_{\zeta}$ yields
\be\label{ht}
\delta \langle H_{\zeta}\rangle=\frac{\Omega_{d-2}l^d}{d^2-1}\delta \langle T_{00}\rangle,
\ee
where $\delta \langle T_{00}\rangle$ is the change in the energy density in comparison to the ground state.
 For non-conformal matter field, it was assumed in \cite{jacobson16} that \be\label{Kva}
\delta \langle E\rangle=\frac{\Omega_{d-2}l^d}{d^2-1}\big(\delta \langle T_{00}\rangle+\delta X \big),
\ee
where $\delta X$ is some scalar in the QFT.
 $X$ was first introduced in \cite{jacobson16}, because if the matter field is not conformal, $E$ is not given by (\ref{hz}) and one cannot use (\ref{ht}). Since the spatial Ricci scalar  centered at $O$ is equal to twice the FNC 00-component of the
 spacetime Einstein tensor
 \be
 \mathcal{R}=2 G_{00}.
 \ee
Assuming $|\delta V| /{(G l \delta t)}=\varsigma \delta E$ and plugging in (\ref{variation}) and (\ref{Kva}) in this relation, we obtain
 \bea \label{derivation}
  \frac{\Omega_{d-2}l^{d}}{3(d^2-1)G} G_{00}
 &=& \varsigma \frac{\Omega_{d-2}l^d}{d^2-1}\left(\delta \langle T_{00}\rangle+\delta X \right).
 \eea
 Combining the above result and (\ref{Kva}) in all reference frame and position, one can  then achieve the Einstein equation
 \be\label{einstein}
 G_{ab}=8\pi G \big(\delta \langle T_{ab}\rangle+\delta X_{ab} \big).
 \ee
 where we have set $\varsigma= 8\pi/3 $. The CV conjecture shows its power given that the growth rate of quantum complexity and the volume changed per unit time are associated to the variation of energy.  Extended the above discussion to the framework of conformal Fermi coordinates, the results obtained here might be applicable to horizon scale (see Appendix \ref{B1} for related discussions).

\subsection{ Friedmann equations of cosmology}
\label{section31}
As a warm-up exercise for the application of the CA duality in cosmology, we shall discuss how to derive the cosmological Friedmann equations in this model.
The standard  Friedmann-Lemaitre-Robertson-Walker (FLRW) metric in Cartesian cooridinates is given by
\be
ds^2=-dt^2+a(t)^2 \frac{d\vec{x}^2}{(1+\frac{1}{4}K \vec{x}^2)^2},
\ee
where $\vec{x}^2=\delta_{ij}x^i x^j$ and $K$ is the space curvature.
The retrad frame associated to a comoving geodesic is
\bea
(e_0)^{\mu}=(1,0,0,0), ~~~(e_1)^{\mu}=a^{-1}(0,1,0,0),\\
(e_2)^{\mu}=a^{-1}(0,0,1,0), ~~~(e_3)^{\mu}=a^{-1}(0,0,0,1).
\eea
The original coordinates is related to the Fermi normal coordinates up to third order in the affine parameter
\be
t=t_{F}-\frac{H\vec{x}^2_{F}}{2},~~~~
x^i=\frac{x^i_{F}}{a(t_{F})}\bigg(1+\frac{H^2 \vec{x}^2_{F}}{3}\bigg),
\ee
where $H$ is the Hubble constant.
The Fermi normal coordinates can be obtained  via the metric tensor transformation $g^{F}_{\mu\nu}=g_{\alpha\beta}\frac{\partial x^{\alpha}}{\partial x^{\mu}_{F}}\frac{\partial x^{\beta}}{\partial x^{\nu}_{F}}$.
This leads to a new presentation of the FLRW metric in Fermi normal coordinates (for $|\vec{x}|\ll H^{-1}$)
\bea
ds^2=-\bigg[1-\left(\dot{H}+{H}^2 \right)\vec{x}^2_{F}\bigg]dt^2_{F}+\bigg[\delta_{ij}-\left(\frac{K}{a^2}+{H}^2\right)\frac{\vec{x}^2_{F}\delta_{ij}-{x}^i_{F}{x}^j_{F}}{3}\bigg]d{x}^i_{F}d{x}^j_{F}.
\eea
(The FNC obtained here can also be evaluated via (\ref{temporary}-\ref{spatial}) as shown in Appendix B).  Notably, the spatial components of the Ricci scalar can be easily evaluated as
\be\label{FNCR}
{}^FR_{ij}^{~~ij}=6\left(\frac{K}{a^2}+{H}(t_F)^2\right),~~~ i,j=1,2,3.
\ee
The FNC in
the cosmological context are only valid on scales that are much smaller than the horizon,
since it is perturbative in terms of $H x^i_F$; if this quantity becomes order one, the perturbative description of the FNC metric breaks down.

Substituting  (\ref{FNCR}) into (\ref{einstein}), we obtain
\be
 3(H^2+\frac{K}{a^2})=8\pi G\big(\delta \langle T_{00}\rangle+\delta X \big). \label{fried}
\ee
We can assume $\delta X=0$ in what follows. As postulated in \cite{jacobson16}, at the zeroth order, the small geodesic ball is in equilibrium and quantum fields are in their vacuum or ground state and the curvature is that of a Minkowski spacetime.
To the first order, we assume that the geodesic ball is  dominated by some matter and energy and choose to model the matter and energy  by a perfect fluid.
The  energy-momentum tensor for a perfect fluid can be written as
\be
\delta \langle  T_{\mu\nu}\rangle=(\rho+p)U_{\mu}U_{\nu}+p g_{\mu\nu},
\ee
where $ \rho$ and $p$ are the energy density and pressure (respectively) as measured in the rest Fermi frame, and U$\scriptstyle \mu$ is the four-velocity of the fluid.  The four-velocity is given by $U^{\mu}=(1,0,0,0)$.  The  energy-momentum tensor can be simply expressed as $ \delta \langle  T^{\mu}_{\nu}\rangle={\rm diag} (-\rho, p, p,p)$.
We can then recast (\ref{fried}) into the standard Friedmann equation
\be\label{friedmann1}
H^2+\frac{K}{a^2}=\frac{8\pi G}{3}\rho.
\ee
Together with the continuity equation of the perfect fluid $ \dot{\rho}=-3H(\rho+P),$
we obtain the second Friedmann equation
\be
\dot{H}-\frac{K}{a^2}=-4\pi G(\rho+P),
\ee
where the dot  denotes the derivative with respect to $t_F$.
We thus obtain the cosmological Friedmann equations  in the causal diamond. In above discussions, we do not use the first law of entanglement entropy and the maximal vacuum
entanglement hypothesis proposed in \cite{jacobson16}.

\section{ Dark energy from  complexity/action duality}
The CA and CV dualities might be two sides of the same coin, although both proposals have their own merits and related study is still at the preliminary stage. Since the growth rate of quantum complexity $\mathcal{C}$ bounded by energy is equivalent to the quantum velocity $d\mathcal{S}/dt$ as shown in (\ref{veloc}), the CA duality can also be interpreted as the $\mathcal{S}$-action duality. Connections built between  cosmology and the CV duality imply that the evolution of our universe may closely relate to quantum complexity. As the underlying origin of dark energy is still unknown, we propose to consider that the cosmic acceleration is driven by the growth rate of quantum complexity.

The Einstein-Hilbert action is given by $\mathcal{A}=\frac{1}{16 \pi G} \int d^4x \sqrt{-g} R$. From the CA duality and the definition of the quantum complexity $\mathcal{C}$, we have
\be \delta \mathcal{C}=\delta \mathcal{A}= \frac{\gamma}{16 \pi G}R \delta V \delta t,\ee
where $\gamma$ is a constant to be determined.
Defining energy density as $\rho=\delta E/\delta V$, from the CA duality and the Einstein-Hilbert action one finds that the energy density behaves as
\be \rho= \frac{\delta\mathcal{C}}{\delta V \delta t}=\frac{\gamma}{16 \pi G}R . \ee
The FLRW metric in spherical coordinates with $K=0$ reads
\be
ds^2=- dt^2+ a(t)^2 \big(dr^2+r^2 d \theta^2+r^2 \sin^2\theta d\phi^2\big).
\ee
The Ricci scalar simply takes the form $R=6\big(\frac{\ddot{a}}{a}+\frac{\dot{a}^2}{a^2}\big)$. We assume the energy density associated to the quantum complexity dominates on  the right hand of the Friedmann equation and then  (\ref{friedmann1}) becomes
\be
H^2=\gamma (\dot{H}+2 H^2).
\ee
This equation can be simply solved by introducing $x=\ln a$ and the Friedmann equation takes a new form
\be \label{maineq}
H^2=\gamma \bigg(\frac{1}{2}\frac{d H^2}{dx}+2 H^2\bigg).
\ee
The solution to (\ref{maineq}) is given by
\be
H^2=c_0 e^{\frac{2x}{\gamma}-4x}=c_0  a^{\frac{2}{\gamma}-4}.
\ee
The energy density thus takes the form $\rho\sim H^2\propto a^{\frac{2}{\gamma}-4}$. Comparing  with the standard literature of cosmology \cite{sean} and the equation of state, one measures the state parameter $w$ as in $\rho\sim a^{-3(1+w)}$. The state parameter $w$ is then
\be
w=\frac{1}{3}-\frac{2}{3 \gamma}.
\ee
 Therefore, from the CA conjecture, we obtain a pattern of energy driving the accelerated expansion of the universe. The accelerating universe requires $w<-1/3$ and this leads to $\gamma<1$. The most recent {\it Planck}  data combining with other astrophysical data indicate $w=-1.006 \pm 0.045$ \cite{Planck16}. If we take $\gamma=1/2$, the state parameter $w=-1$ is similar to the cosmological constant.
As $0<\gamma< 1/2$,  $w<-1$, a value same as the phantom model \cite{phantom}.

\section{Conclusion and discussion}
In summary, as quantum complexity can be measured by the shortest curve of geodesic quantum distance, we have studied the quantum complexity in a new framework related to the emergence of Einstein's equation by exploring the hidden connections between concepts of quantum information theory and the geometry of gravity. We have also derived the Friedmann equations from the CV duality by taking the FLRW geometry as small corrections to the local Minkowski spacetime in FNC system. The derivation presented here is comparable to the derivation of Friedmann equations from thermodynamics involving apparent horizons of cosmology previously given in \cite{cai1,cai2,solidspace}.

 The accelerated expansion of the universe can be interpreted as driven by the quantum complexity if the  spacetime geometry can be viewed as an entanglement structure of the microscopic quantum state. It appears that the complexity driven dark energy scenario is able to avoid the fine tunning problem because the energy density is not associated to high energy scales up to the Planck scale. Since the dark energy is proportional to the Ricci scalar, it should be relatively small compared to the Hubble ratio during radiation dominated era and become comparable to non-relativistic matter in the matter dominated era. This may in the end solve the coincidence problem. In the forthcoming work \cite{complex17}, we will carefully examine the evolution and structure formation of the universe in this model. Further crosschecks including comparisons between our proposal and other existing dark energy models (for example \cite{verlinde,mli,mli2, bwang,bwang2}) will also be discussed more carefully in our follow up paper.
 \\

 \textit{\textbf{Note added}}: While finalizing this work, we received the paper \cite{chapman17} defining complexity $\mathcal{C}$ of quantum field theory states as the minimal length from a reference state to a target state calculated via the Fubini-Study metric in the quantum field theory. But the authors did not study the relations  between complexity and cosmology.

\section*{Acknowledgement}
 We would like to thank Ted Jacobson, John McGreevy, Yu Tian and Shao-Feng Wu for helpful discussions at the early stage of this work.
 XHG was partially supported by NSFC, China (No.11375110).  BW was partially supported by NSFC
grants (No.11575109).

\appendix
\section{Quantum Fisher information metric and relative entropy }

The quantum Fisher information metric can also be derived from the concept of \textit{relative entropy}. In quantum information theory,  relative entropy is a measure of distinguishibility between a state $\rho$ and a reference state $\sigma$ associated with the same Hilbert space $\mathcal{H}$. It is defined as
\be
S(\rho\parallel \sigma)={\rm tr}(\rho \log \rho)-{\rm tr }(\rho \log \sigma).
\ee
The quantum   relative entropy is a ¡°mother quantity¡± for other entropies in quantum information theory, such as the quantum entropy,
the conditional quantum entropy, the quantum mutual information, and the conditional quantum mutual information. So we can re-express many of the
entropies in terms of relative entropy. For example, the mutual information of two disjoint subsystems $A$ and $B$ is given by
\be
I(A:B)=S(A)+S(B)-S(A\cup B),
\ee
where $S(A)=-{\rm Tr}{\rho_A \log \rho_A}$ and $S(A\cup B)=-{\rm Tr}{\rho_{AB} \log \rho_{AB}}$.
One can prove that
\bea
I(A:B)=S(\rho_{AB}\parallel \rho_A \otimes \rho_B)
={\rm Tr}{\rho_{AB}[\log \rho_{AB}-\log(\rho_{A}\otimes \rho_{B})]}.
\eea
So does the  conditional quantum entropy  $ S(A|B)\equiv S(A\cup B)-S(B)$, that is to say $S(A|B)=-S(\rho_{AB}\parallel I_A \otimes \rho_B)$. The positiveness of relative entropy then infers the non-negativity of quantum mutual information and negativity of conditional quantum entropy. One of the most remarkable properties of quantum entropy and  a radical departure from the intuitive classical entropy is that one can sometimes be more certain about the joint state of a quantum system than we can be about any one of its individual parts.  This is the fundamental reason that conditional quantum entropy can be negative.

The quantum Fisher information metric can be obtained from relative entropy by considering
\bea
S(\sigma+\epsilon \sigma \parallel \sigma )=S(\sigma+\epsilon \sigma \parallel \sigma)\big|_{\epsilon=0}+\epsilon S'(\epsilon)\big|_{\epsilon=0}+\frac{1}{2}\epsilon^2 S''(\epsilon)\big|_{\epsilon=0}.
\eea
One finds $S(0)=S'(0)=0$ and $S''(\epsilon)$ is the term related to the Fisher information metric.
In general, a set of probability distribution $p_{\theta}=p(\theta)$ parameterized by $\theta^i$ with $i\in {1,...,n}$ is a manifold. The Riemanian metric on this manifold is Fisher information metric defined in an integral form
\be
\mathcal{G}_{ab}= \int_X p_{\theta}\bigg(\frac{1}{p_{\theta}}\frac{\partial p_{\theta}}{\partial \theta^{a}}\bigg)\bigg(\frac{1}{p_{\theta}}\frac{\partial p_{\theta}}{\partial \theta^{b}}\bigg)d^4 X .
\ee
\section{ Brief review on Fermi Normal Coordinates}
\begin{figure}
\begin{center}
	\includegraphics[height=7cm,trim={0cm  19cm 0cm 3.5cm},clip]{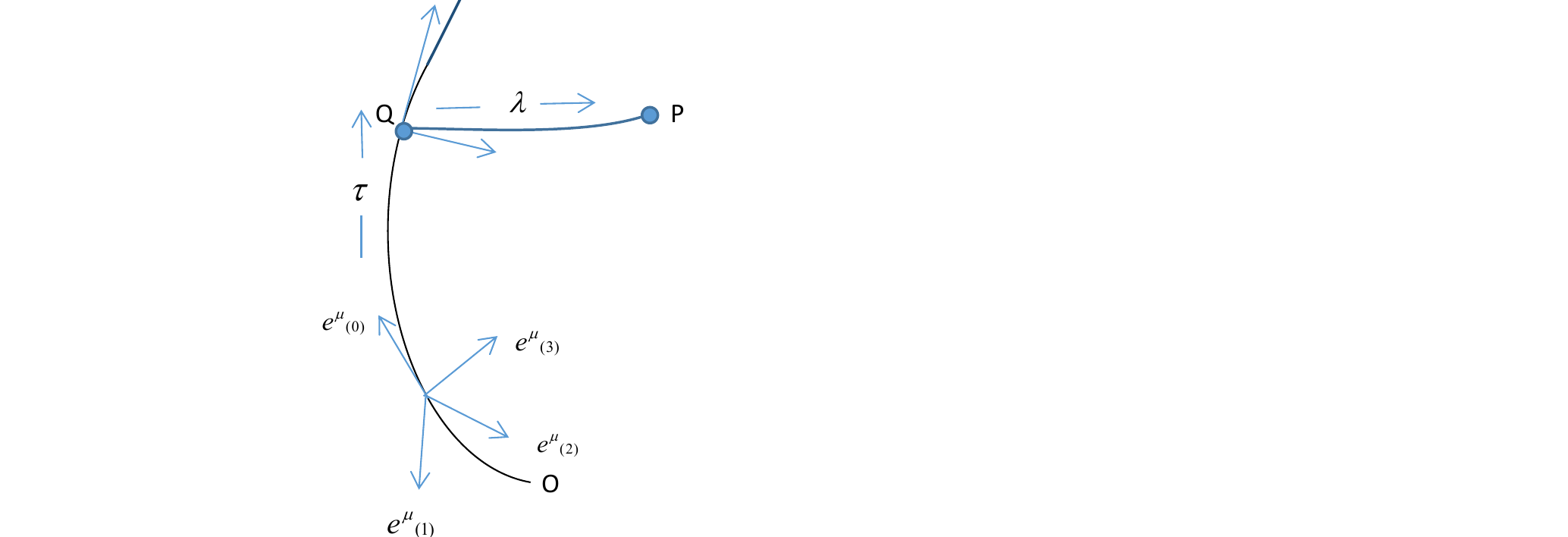}
\end{center}
	\caption{Geometrical construction of the Fermi normal coordinates. Fermi normal coordinates are determined by an orthonormal reference frame $e^{\mu}_{(\alpha)}$.   $e^{\mu}_{(0)}$ is the timelike unit vector that is tangent to the central geodesic $O$. The point $P(x^{\mu}_F)$ is found by first following $O$ for a proper time $\tau$ and following a certain orthogonal geodesic at a proper distance $\lambda$.  }
\end{figure}
The  Equivalence Principle of General Relativity asserts that, in the presence of a gravitational field, the physical laws of the inertial reference frame are valid in an infinitesimally small laboratory. At each event in spacetime, the spacetime is locally flat.  Therefore, it is
possible to introduce Riemann normal coordinates that constitute a geodesic system of coordinates that is inertial at the event under consideration.  The basic idea behind Riemann normal coordinates is to use the geodesics through a given point to define the coordinates for nearby points. However, for the physical interpretation of measurements by a free observer, the Riemann normal coordinate system is no longer very suitable. One then calls for the Fermi normal coordinates,  which is a normal geodesic coordinate system in a cylindrical region
about the worldline of the observer \cite{Fermi}. The Fermi normal coordinates provide a powerful tool in which a freely falling observer can report observations and local experiments. As a natural extension of the Riemann normal coordinates, the Fermi normal coordinates is valid for a limited region of space and for \textit{all time} \cite{Fermi}.

Let us see how the Fermi normal coordinates are constructed from a geometric point of view. Imagine  a free falling observer moving along a worldline $(\bar{t}, \bar{x}^i)$, a spacetime region with coordinates $x^{\mu}=(t,x^i)$. The free falling observer carries an orthonormal parallel-propagated tetrad frame ${e}^{\mu}_{(\alpha)}$ along its path such that the temporal component ${e}^{\mu}_{(0)}=d x^{\mu}/d\tau$ with $\tau$ the proper time along the worldline of the observer $O$.  This means that ${e}^{\mu}_{(0)}$ is the timelike unit tangent vector of the worldline of $O$ and behaves as the local temporal axis. Meanwhile ${e}^{\mu}_{(i)}$ ($i=1,2,3$) are orthogonal spacelike unit axes forming the local spatial frame of the observer. A local hypersurface can be constructed via  the class of spacelike geodesics orthogonal
to the worldline at each event $Q(\tau)$. Consider a point $P$ in the vicinity of  $Q(\tau)$ with coordinates $x^{\mu}$ on this hypersurface. There will  be a unique spacelike geodesic connected from $Q(\tau)$ to $P$. We can thus define the Fermi normal coordinates of $Q$ and $P$ to be $ x^{\mu}_{F}=(\tau,\textbf{0})$ and $ {x}^{\mu}_{F}=(t_{F},x^i_{F})$, respectively, satisfying the relation
\be
 \frac{d x^{\mu}}{d \lambda}\bigg|_{\lambda=0}=x^{i}_{F}e^{\mu}_{(i)}
\ee
 where $\lambda$ is the proper
length of this segment from $Q$ to $P$ as in Figure 1. Thus the reference observer $O$ is always located
at the spatial origin of the Fermi normal coordinates. Since in Fermi normal coordinates the metric is rectangular on $O$, the spacetime at $O$ yields
\be
g_{\mu\nu}\mid_O\equiv\eta_{\mu\nu}, ~~~\Gamma^{\mu}_{\rho\nu}\mid_O=0.
\ee
Under this construction, the coordinate transformation from some coordinate system (for example the Schwarzschild metric ) $x^{\mu}$ to the Fermi normal coordinate can be computed order-by-order in $x^i_{F}$ by repeatedly using the  geodesic
equation. In detail, we can construct the mapping between arbitrary coordinates $x^{\mu}$ and the Fermi coordinates defining the geodesic.
This can be realized by solving the geodesic equation
\be
\frac{d^2 x^{\nu}}{d \lambda^2}+\Gamma^{\nu}_{\rho\beta}\frac{dx^{\rho}dx^{\beta}}{d \lambda d \lambda}=0,
\ee
which can be solved perturbatively using the power law expansion of $x^{\mu}$
\be
x^{\mu}=\alpha^{\mu}_{0}+\alpha^{\mu}_{1}\lambda+\alpha^{\mu}_{2}\lambda^2+....
\ee
where
\bea
\alpha^{\mu}_{0}&=&(t_F,0,0,0),\\
\alpha^{\mu}_{1}&=&\frac{d x^{\mu}}{d \lambda}\bigg|_{\lambda=0}=x^i_{F}(e_i)^{\mu},\\
\alpha^{\mu}_{2}&=&\frac{1}{2!}\frac{d^2 x^{\mu}}{d \lambda^2}\bigg|_{\lambda=0}=-\frac{1}{2}\Gamma^{\mu}_{\gamma\nu}\alpha^{\gamma}_{1}\alpha^{\nu}_{1}.
\eea
In brief, the spacetime elements expanded in Taylor series in terms of Fermi normal coordinates can be written as $g_{\mu\nu}=\eta_{\mu\nu}+h_{\mu\nu}(x^{i}_{F})$, namely
\bea
g_{00}&=&-1-{}^FR_{0i0j}(O)x^{i}_{F}x^{j}_{F}+\cdots,\label{temporary}\\
g_{0i}&=&-\frac{2}{3}{}^FR_{0jik}(O)x^{i}_{F}x^{k}_{F}+\cdots,\\
g_{ij}&=& \delta_{ij}-\frac{1}{3}{}^FR_{ikjl}(O)x^{k}_{F}x^{l}_{F}+\cdots,\label{spatial}
\eea
where  ${}^FR_{ikjl}$ denotes the projection of the Riemann tensor on the observer¡¯s tetrad along the reference trajectory under the relation
\be\label{riemann}
{}^FR_{\gamma\kappa\rho\sigma}=R_{\mu\nu\alpha\beta}e^{\mu}_{(\gamma)}e^{\nu}_{(\kappa)}e^{\alpha}_{(\rho)}e^{\beta}_{(\sigma)}.
\ee
Note that  $R_{\mu\nu\alpha\beta}$ denotes the curvature components with respect to the global frame.
The validity of this expansion is closely related to the validity of the Fermi coordinates and the curvature of the spacetime.

\subsection{Conformal Fermi Coordinates and Friedmann Equations}\label{B1}
Although Fermi normal coordinates are a useful frame for the local observers, the FNC are only valid on scales much smaller than the cosmological horizon. For the purpose of cosmological application, the authors in \cite{CFC2015} introduced the so-called conformal Fermi coordinates (CFC), which not only preserve all the advantages of FNC, but also are valid outside the horizon. The CFC are constructed in the vicinity of a timelike central geodesic which are the  same as FNC. However, we  will not restrict the local spacetime to be rigorously Minkowski, but the ``conformal Minkowski spacetime" allowing for a homogeneous expansion over time.
Namely, in the CFC frame, the lowest order CFC metric is a flat FLRW spacetime. The CFC metric thus takes the following form
\be\label{cfcmetric}
g^F_{\mu\nu}(x^{\mu}_F)=a^2(\tau_F)\left[-\eta_{\mu\nu}+\mathcal{O}[(x^i_F)^2]\right],
\ee
where $\mathcal{O}[(x^i_F)^2]$ denotes corrections to the conformally flat part starting at the quadratic order
in $x^i_F$. Note that the CFC time $a(\tau_F)$ should be some
suitable conformal time rather than the observer¡¯s proper time. For simplicity, we would like to introduce the conformal metric $\tilde{g}_{\mu\nu}\equiv a^{-2}(\tau_F)g_{\mu\nu}$.

The construction of CFC is given as follows:

 a). Firstly, we choose the same set of orthogonal tetrad $e^{\mu}_{(\alpha)}$ as in the construction of FNC. The proper time $t_{F}$ characterizes the observer's geodesic in the usual way. Secondly, we consider a positive spacetime scale $a_{F}$ at a point along the central geodesic at the proper time $t_F$. Define a ``conformal proper time" $\tau_F$ as our time coordinate
\be
d \tau_F=a^{-1}_F\big(P(t_F)\big) d t_{F}.
\ee
The point $P$ has CFC coordinates $(\tau_F, \textbf{0})$.

b). Consider a family of \textit{conformal geodesic } $\tilde{h}(\tau_F;\alpha^i;\lambda)$ with respect to $\tilde{g}_{\mu\nu}$ with the affine parameter at $P$ given by $\lambda=0$ and the tangent vector at $P$ given by $\alpha^i e^{\mu}_{(i)}$, with $\alpha^i$ constants specifying the initial direction of the geodesic and $\lambda$ measures the geodesic distance with respect to the conformal metric \cite{CFC2015}.

c). The point $Q$ with coordinates $(\tau_F, x^i_{F})$ is located on the conformal geodesic $\tilde{h}(\tau_F;a_F(P)\beta^i;\lambda)$ where $\lambda=(\delta_{ij}x^i_{F}x^j_{F})^{1/2}$
and $\beta^i=x^i_F/\lambda$ \cite{CFC2015}. This guarantees the proper distance squared form $P$ to $Q$ is $a^2_{F}\delta_{ij}x^i_{F}x^j_{F}$ at the lowest order, which is exactly depicted in the metric (\ref{cfcmetric}).

To the quadratic order, the CFC metric can be related to the conformal Riemann curvature tensor through
\bea
g^F_{00}(x_F)&=&a^2_{F}(\tau_F)\bigg[-1-{}^F\tilde{R}_{0i0j}(T)x^{i}_{F}x^{j}_{F}\bigg],\\
g^F_{0i}(x_F)&=&a^2_{F}(\tau_F)\bigg[-\frac{2}{3}{}^F\tilde{R}_{0jik}(T)x^{i}_{F}x^{k}_{F}\bigg],\\
g^F_{ij}(x_F)&=&a^2_{F}(\tau_F)\bigg[\delta_{ij}-\frac{1}{3}{}^F\tilde{R}_{ikjl}(T)x^{k}_{F}x^{l}_{F}\bigg],
\eea
where ${}^F\tilde{R}_{ikjl}$ is the Riemann curvature tensor constructed with respect to $\tilde{g}_{\mu\nu}$ in the CFC frame.  Similar as (\ref{riemann}), the Riemann tensor in CFC frame can be expressed in global coordinates
\be
{}^F\tilde{R}_{\gamma\kappa\rho\sigma}=\tilde{R}_{\mu\nu\alpha\beta}\tilde{e}^{\mu}_{(\gamma)}\tilde{e}^{\nu}_{(\kappa)}\tilde{e}^{\alpha}_{(\rho)}\tilde{e}^{\beta}_{(\sigma)},
\ee
where $\tilde{R}_{\mu\nu\alpha\beta}$ is the Riemann tensor of the conformal metric $\tilde{g}_{\mu\nu}$ computed in the global coordinates. For the flat universe with $k=0$, the spatial component of the Ricci scalar is $\mathcal{R}=6 \dot{a}^2/a^2$.

Similar to the section \ref{section3}, we consider a geodesic causal ball with radius $l$ without the limitation that $l$ must be smaller than the local curvature length.  Repeating the procedure given in section \ref{section31}, we can  also obtain the Friedmann equations in the CFC frame for flat universe.
 The derivation is not restricted to a small ball size.

\end{document}